\documentclass
[twocolumn,showpacs,10pt,notitlepage,tightenlines,lengthcheck,superscriptaddress,prl]{revtex4}%
\usepackage{amsfonts}
\usepackage{amsmath}
\usepackage{amssymb}
\usepackage{graphicx}%
\providecommand{\U}[1]{\protect\rule{.1in}{.1in}}

\begin{document}
\title{Photo-Thermoelectric Effect at a Graphene Interface Junction}
\author{Xiaodong Xu}
\affiliation{Center For Nanoscale Systems}
\author{Nathaniel M. Gabor}
\author{Jonathan S. Alden}
\affiliation{Center For Nanoscale Systems} \affiliation{Laboratory
of Atomic and Solid State Physics, Cornell University, Ithaca, New
York 14853 }
\author{Arend van der Zande}
\affiliation{Laboratory of Atomic and Solid State Physics, Cornell
University, Ithaca, New York 14853 }
\author{Paul L. McEuen}
\email{mceuen@ccmr.cornell.edu} \affiliation{Center For Nanoscale
Systems} \affiliation{Laboratory of Atomic and Solid State Physics,
Cornell University, Ithaca, New York 14853 } \keywords{one two
three}\
\begin{abstract}
We investigate the optoelectronic response of a graphene interface
junction, formed with bilayer and single-layer graphene,  by
photocurrent (PC) microscopy. We measure the polarity and amplitude
of the PC while varying the Fermi level by tuning a gate voltage.
These measurements show that the generation of PC is by a
photo-thermoelectric effect. The PC displays a factor of $\sim$10
increase at the cryogenic temperature as compared to room
temperature. Assuming the thermoelectric power has a linear
dependence on the temperature, the inferred graphene thermal
conductivity from temperature dependent measurements has a $T^{1.5}$
dependence below $\sim$100 K, which agrees with recent theoretical
predictions.

\end{abstract}
\pacs{78.20.Nv, 73.23.-b, 72.40.w, 73.50.Lw} \maketitle

When a photosensitive device is illuminated by light, an electric
current, known as photocurrent (PC), can be generated. PC generation
in semiconductor optoelectronic devices is mainly due to separation
of the excited electron and hole pair by a built-in electric field,
as shown in Fig. 1(a). It is also known that if a temperature
gradient is generated by light across an interface between two
materials, which have different thermoelectric power ($S$), there is
PC generation by the photo-thermoelectric effect (PTE), as shown by
Fig. 1(b). The magnitude of the generated PC is directly
proportional to $S$, which is also a measure of the partial molar
entropy. Since entropy is proportional to the density of states
(\textit{D(E)}), from the second law of thermodynamics, the hot
carriers tend to diffuse to the material with larger \textit{D(E)}
to maximize the entropy.

Graphene is an interesting material with unusual electronic,
optical, and thermal properties ~\cite{GeimGrapheneFab,GeimQH, Chen,
LauNanoletter,Yzhang,KimQH}. Since the conduction and valence band
touch each other at the Dirac
point~\cite{GeimGrapheneFab,KimQH,Semenoff}, there is no bandgap for
graphene. A question naturally arises as to which mechanism
dominates the PC generation in graphene optoelectronic devices.
There have been a few studies focusing on the room temperature
opto-electronic response at the junction formed by single layer
graphene and metal
contacts~\cite{GiovannettiPRL,LeeNatureNano,AvourisNanoLetter,ParkNanoletter,AvourisPRB}.
The generated PC is interpreted based on the built-in electric field
picture. However, recent transport measurements demonstrated the
thermoelectric effect in graphene transistor
devices~\cite{KimGrapheneThermal,LauGraphenePRL}, which suggests the
PTE may play an important role in PC generation in graphene devices.
It is the aim of this paper to elucidate the fundamental physical
mechanism giving rise to the opto-electronic response at
zero-bandgap graphene heterostructures.

\begin{figure}[ptb]
\centerline{ \scalebox{.80} {\includegraphics{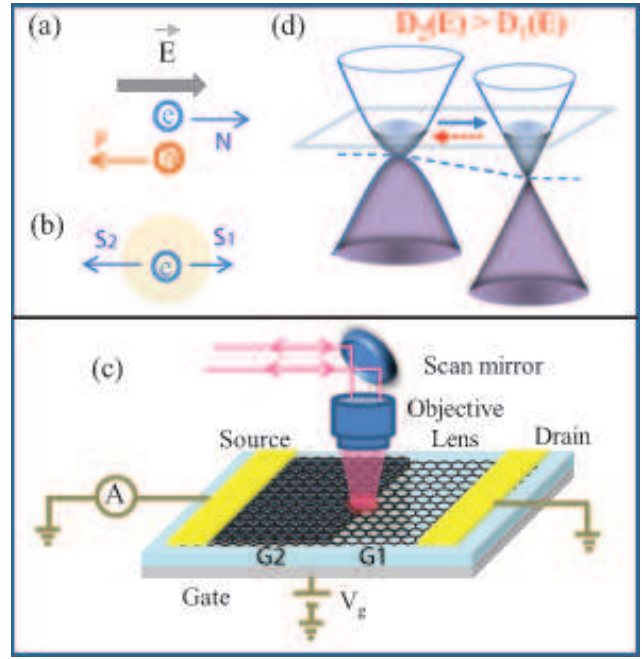}}}
\caption{(color online) (a) The built-in electric field picture for
PC generation. The direction of the field is defined along the
direction of electron movement. (b) Hot carrier diffusions. (c)
Schematics of the experimental setup and device geometry. (d)
Aligned Fermi level of bilayer (left) and single layer (right)
graphene. The blue and red dashed arrow represent the electron flow
direction induced by the built-in electric field and by the
thermoelectric
effect, respectively. }%
\label{fig. 1}%
\end{figure}

A graphene interface junction, formed by single and bilayer graphene
(G1/G2) as shown in Fig. 1(c), will give rise to opposite signs of
PC for the two different mechanisms. Thus it provides a unique
opportunity to identify the origin of PC. Since single-layer
graphene has a linear energy-momentum dispersion relation and
bilayer has a quadratic dispersion relation, the \textit{D(E)} of
single-layer ($D_{1}(E)\propto E$) is smaller than that of bilayer
($D_{2}(E)\propto$ finite constant) when the Fermi energy ($E_{f}$)
is not far away from the Dirac point. Thus, for the same charge
density, $|E_{f}|$ of a single-layer is larger than that of the
bilayer, i.e. there is a built-in potential difference.

\begin{figure}[ptb]
\centerline{ \scalebox{.52} {\includegraphics{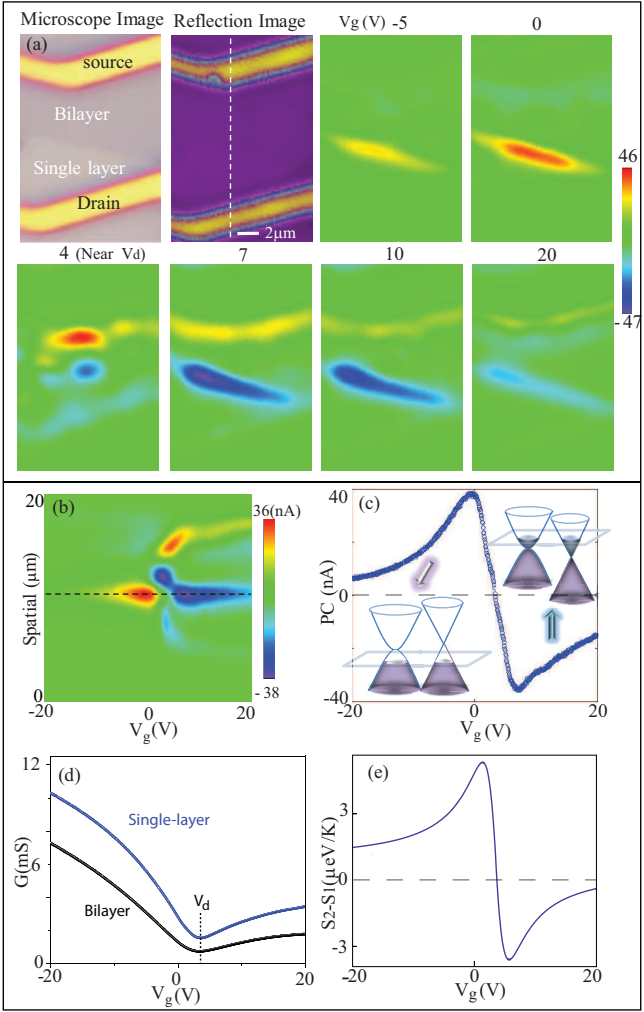}}}
\caption{(color online) Data are taken at T=12K. (a) PC images at
various $V_{g}$. (b) The PC image obtained by the laser linescan as
a function of $V_{g}$. The laser scan position is indicated by the
dashed white line in the reflection image. (c) PC response at the
center of G1/G2 as a function of gate voltage. The top right (bottom
left) inset is the aligned Fermi level between the single and
bilayer at the \textit{n (p)} doping. (d) Conductance measurement of
single (blue) and bilayer (black) graphene as a function of $V_{g}$.
(e) Calculated thermoelectric power
difference at G1/G2 as a function of $V_{g}$.}%
\label{fig. 2}%
\end{figure}

The aligned Fermi level between single and bilayer graphene leads to
the Dirac point of the single layer being lower than the bilayer, as
shown in Fig. 1(d). According to the electric field picture, the
photo-excited electrons would be expected to flow from the bilayer
to the single-layer, shown by the blue arrow, and result in a
positive PC in the present experimental setup. However, if the PTE
is the correct mechanism generating PC, since $D_{2}(E)>D_{1}(E)$,
the electrons should flow from the single to the bilayer and result
in a negative PC, shown by the red dashed arrow. By identifying the
sign of the PC experimentally, we can determine which mechanism
dominates.

The graphene device is fabricated by mechanical exfoliation of
graphite sheets onto a 90nm $SiO_{2}/Si$ wafer~\cite{GeimQH}. The
single and bilayer graphene are identified by optical contrast and
Raman spectroscopy~\cite{GeimRamanPRL}. Au/Cr or Au/Ti electrodes
are deposited using photolithographic patterning or shadow mask
techniques. The device is held in a vacuum cryostat with a
temperature control from $\sim$ 10 to 300 K. The PC and the
correlated reflection image are simultaneously obtained by scanning
the laser across the
device~\cite{LeeNatureNano,AvourisNanoLetter,ParkNanoletter,AvourisPRB}.
The laser excitation wavelength is fixed at 635 nm and the laser
spot is about 1 $\mu m$. All the PC images presented in this work
are taken at zero source-drain bias. We have measured eight
different devices and obtained consistent results.

Figure 2(a) displays the PC images of device 1 with the same color
scale at various gate voltages ($V_{g}$) and a temperature of 12 K.
Pronounced PC is seen at the graphene-metal contact interface (G/M)
and G1/G2. The PC image in Fig. 2(b) is taken by scanning the laser
along the dashed white line indicated in the refection image while
sweeping the gate voltage continuously. The PC generation at the G/M
has been intensively studied and was mainly attributed to carrier
separation by the built-in electric field
\cite{LeeNatureNano,AvourisNanoLetter,ParkNanoletter,AvourisPRB}. In
the following, we will focus on the optoelectronic response from
G1/G2.

By tuning the gate voltage $V_{g}$ from smaller than $V_{d}$ to
larger than $V_{d}$, where $V_{d}$ corresponds to the Dirac point as
shown in Fig. 2(d), the majority carrier in the graphene changes
from hole to electron. The PC at G1/G2 switches signs, changing from
positive (red) to negative (blue). Figure 2(c) shows the gate
voltage dependence of the PC at G1/G2, which is the linecut across
the dashed black line in Fig. 2(b). The PC amplitude evolves as the
gate voltage varies. On the hole doping side, the PC amplitude
increases first, reaches a maximum, and then decreases as the gate
voltage increases. The same observation holds for electron doping.

With all the above experimental observations, we determine that the
PTE dominates the PC generation at G1/G2, instead of the built-in
electric field. Our conclusion is primarily based on the fact that
dominance by the built-in electric field would result in positive
(negative) PC for the electron (hole) doping, which is clearly
opposite to the experimental observations.

The physical picture for PC generation due to the PTE is the
following: after the electrons are excited from the valence band to
the conduction band, they initially relax back to the Fermi level on
the time scale of $\sim fs$ by phonon emission and form a hot
Fermion distribution~\cite{SundongPRL,RanaNanoLetter}. Since
$D_{2}(E)$ is larger than $D_{1}(E)$, the hot free carriers tend to
diffuse from the single-layer into the bilayer due to the
temperature gradient across G1/G2, which leads to a negative
(positive) current for electron (hole) doped graphene.

To make a quantitative comparison between the theory and experiment,
the PC generated by the PTE can be formulated as
\begin{equation}
I=\frac{(S_{2}-S_{1})\times\Delta T}{R},
\end{equation}
where $S$ is thermoelectric power, $R$ is the resistance, and
$\Delta T$ is the temperature difference. From the Mott
relation~\cite{MottPR,KimGrapheneThermal,LauGraphenePRL}, we have
the Seebeck coefficient as
\begin{equation}
S=-\frac{\pi^{2}k_{b}^{2}T}{3e}\frac{1}{G}
\frac{dG}{dE}\mid_{E=E_{f}}
\end{equation}
where $k_{b}$ is the Boltzmann constant, \textit{e} is electron
charge, \textit{T} is temperature, \textit{G} is conductance, and
$E_{f}$ is the Fermi energy. The conductance \textit{G} is
proportional to $ne\mu$ for graphene, where \textit{n} is charge
density and $\mu$ is the electron mobility. When $E_{f}$ is away
from the Dirac point, $\mu$ is approximately a constant and $S$ is
proportional to \textit{D(E)}.

\begin{figure}[ptb]
\centerline{ \scalebox{.58} {\includegraphics{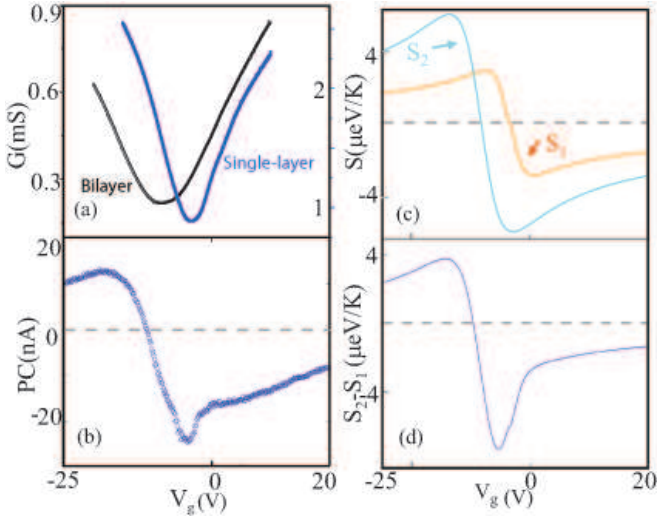}}}
\caption{(color online)  (a) The bilayer (black line) Dirac point is
about 5 V smaller than that of the single-layer (blue line). (b) The
generated PC at the G1/G2 as a function of $V_{g}$. (c) The
calculated $S$ and (d) $S_{2}-S_{1}$.}
\label{fig. 4}%
\end{figure}

The calculated $S_{2}-S_{1}$ as a function of $V_{g}$ is shown in
Fig. 2(e). In the calculation, the $\frac{1}{G} \frac{dG}{dE}$ term
is replaced by $\frac{1}{G} \frac{dG}{dV_{g}}\frac{dV_{g}}{dE}$,
where $\frac{dG}{dV_{g}}$ can be derived from the conductance
measurements. The dependence of $E_{f}$ on the charge density
\textit{n} can be derived from tight binding
calculations~\cite{FermiEnergyvsChargeDensity}. For single layer
graphene, $E_{f}=\hbar v_{F} \sqrt{\pi n}$ and $v_{F}$ is the Fermi
velocity. For bilayer, $E_{f}=\frac{1}{2}\sqrt{(2\hbar v_{F})^{2}\pi
n+2 \gamma_{1}^{2}-2 \gamma_{1} \sqrt{(2\hbar v_{F})^{2}\pi n+
\gamma_{1}^{2}}}$ and $\gamma_{1}$ is the interlayer coupling
strength. The calculated $S_{2}-S_{1}$ qualitatively reproduce the
lineshape and sign of the experimental data in Fig.
2(c)~\cite{note}. $S_{2}-S_{1}$ reverses sign at $V_{g}=V_{d}=$
3.7V, which is close to the sign switch of PC at 3.4V.

The PTE can account for the experimental results from devices with
non-overlapping Dirac points between the single and bilayer. An
example (device 2) is shown by the gate dependent conductance
measurement in Fig. 3(a), where the bilayer Dirac point is $\sim$5V
lower than that of the single-layer. The PC at the G1/G2 as a
function of $V_{g}$ is shown in Fig. 3(b). When $V_{g}$ is between
the two Dirac points, i.e. the single-layer is \textit{p} doped but
the bilayer is \textit{n} doped, $S$ of the single-layer has the
opposite sign of bilayer, as shown in Fig. 3(c). Thus, the
difference in $S$ reaches a maximum at a certain $V_{g}$ between the
two Dirac points, which corresponds to a maximum in the PC data. The
calculated $S_{2}-S_{1}$ for device 2, shown in Fig. 3(d),
qualitatively reproduces the lineshape and sign of the PC.

We also performed temperature and power dependent studies of PC. We
plot the absolute PC amplitude of device 2 at $V_{g}=-4$V as a
function of temperature in Fig. 4(a). The PC amplitude decreases
nonlinearly as the temperature increases. We replot the data on a
logarithmic scale in Fig. 4(b), which can be separated into two
regions around a temperature of 100 K. The data are fitted with a
line with a slope of -0.5 (-1.5) for below (above) $\sim$100K. For
instance, the PC images of device 2 at 14 K and at 295 K are also
displayed on the left and right of Fig. 4(c).

The PTE naturally explains the temperature dependent data. Equation
(1) shows that the PC is proportional to $(S_{2}-S_{1})/\kappa$,
where $\kappa$ is the thermal conductivity. Since $S$ has an
approximate $T^{1}$
dependence~\cite{KimGrapheneThermal,LauGraphenePRL} and $\kappa$ has
a power law dependence of $T^{\beta}$ with
$\beta>1$~\cite{ThermalConductivity,TM2,Thermal3,ThermalC4}, PC is
expected to have a nonlinear dependence of $1/T^{\beta-1}$, which
agrees with the experimental results. Since the slope in Fig. 4(b)
corresponds to $1-\beta$, we infer that $\kappa$ has a $T^{1.5}$
($T^{2.5}$) dependence below (above) $\sim$100 K.

\begin{figure}[ptb]
\centerline{ \scalebox{.50} {\includegraphics{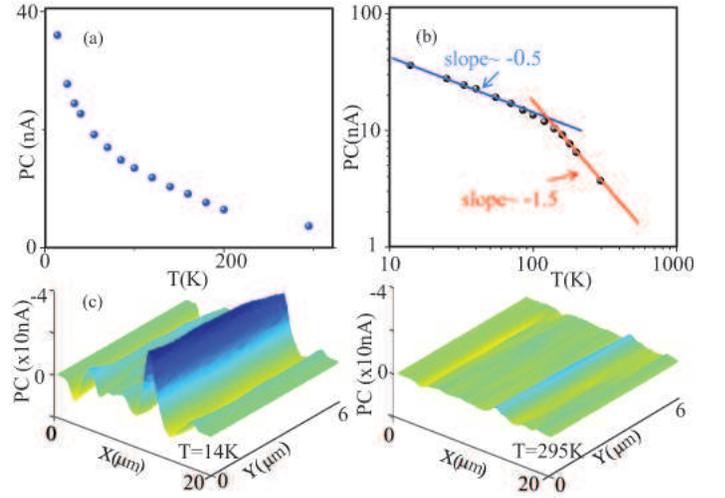}}}
\caption{(color online) (a) The amplitude of PC generated at G1/G2
as a function of T. (b) Logarithmic plot of the data in (a). (c) The
PC images at a temperature of 14 K (left)
and 295 K (right).}%
\label{fig. 4}%
\end{figure}

The $T^{2.5}$ dependence at high temperature is similar to $\kappa$
of the graphite~\cite{ThermalConductivity}. The $T^{1.5}$ dependence
at low temperature agrees with the recent theoretical prediction of
graphene $\kappa$~\cite{ThermalConductivity,TM2,Thermal3,ThermalC4}.
It suggests that at low temperature, the out-of-plane acoustic
phonon mode, which has a quadratic dispersion relation, contributes
to the thermal conductivity. The observation also indicates that the
phonon-induced $\kappa$ dominates the electron-induced $\kappa$ when
$V_{g}$ is close to $V_{d}$~\cite{TM2}.

\begin{figure}[ptb]
\centerline{ \scalebox{.5} {\includegraphics{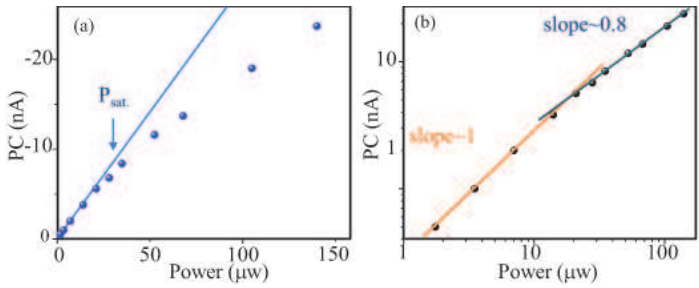}}}
\caption{(color online)  Power dependent PC amplitude generated at
G1/G2 at a temperature of 30K. (a) Linear and (b) logarithmic plots.
Saturation power is defined as the power corresponding to the PC
deviating 10\% from the linear region. }
\label{fig. 5}%
\end{figure}

We can estimate the magnitude of the PC generated by the PTE using
Eq. (1) and (2). $\kappa$ of single-layer graphene has been reported
as $5\times10^{3}W/m\cdot K$ at room
temperature~\cite{LauNanoletter}. Taking the heat flow as a radial
wave, given that $\kappa 2 \pi h \Delta T=P\alpha $, where
\textit{h} is the thickness of graphene of $\sim 3{\AA}$, \textit{P}
is incident laser power of 40 $\mu W$, and $\alpha$ is the
absorption coefficient of $2.3\%$, we infer that $\Delta T$ is on
the order of $\sim$0.1K. Taking $(S_{2}-S_{1})$ on the order of $100
\mu eV/K$ by calculation from Eq . (1) and the resistance of
graphene on the order of $5 K\Omega/\mu m^{2}$, the PC is on the
order of $\sim$2 nA, which is consistent with the experimental
observations at room temperature.

With the knowledge of $\kappa$ and $S$ as functions of \textit{T},
we should be able to predict the power dependence of the PC at low
temperature. From Eq. (1) and (2), we have $I\propto T \Delta T$.
When the laser power is strong enough, the induced temperature
difference $\Delta T$ dominates. Thus, $\Delta T$ can be
approximately taken as $T$, which leads to $I\propto T^{2}$. On the
other hand, from $\kappa \Delta T\propto P$ and $\kappa\propto
T^{\beta}$, we have $T\propto P^{\frac {1}{\beta+1}}$, which leads
to $I\propto P^{\frac {2}{\beta+1}}$. Taking $\beta=1.5$ from the
temperature dependent measurement, we expect that the PC should have
$P^{0.8}$ dependence for strong laser power at low temperature.

The obtained laser-power-dependent PC measurements confirm the above
predictions. Taking the PC amplitude as a function of laser power at
T=30K as an example, shown in Fig. 5(a), we observe a PC saturation
effect. The blue line is a guide to the eye and the PC deviates from
the linear dependence around 20 $\mu W$ (corresponding to $\sim 0.7
KW/cm^{2}$). The data are replotted on a logarithmic scale in Fig.
5(b). A line fit with a slope of $\sim$1 corresponds to the data in
the linear region. With strong laser power, the PC shows a $P^{0.8}$
dependence, which is in excellent agreement with the prediction from
the PTE picture.

A final question is whether the PC at a G/M or the interface P-N
junction is also generated by the PTE. We know that $S$ is negative
for electrons and positive for holes in graphene. For the G/M, the
thermal voltage drives electrons (holes) from the metal contact to
the graphene for the \textit{n (p)} doping, which leads to positive
(negative) PC at the source and negative (positive) at the drain.
The PC reverses polarity as the majority carrier changes from
electron to hole. For the P-N junctions formed inside the graphene,
the hot electrons also diffuse from the P to the N channel. The
above explanations are consistent with the experimental observations
at the G/M and PN junctions. Furthermore, we have done other
measurements (data not shown) on the PC generation at G/M and PN
junctions. The results show similar features as G1/G2, such as
comparable PC amplitude, PC saturation at low temperatures, and
similar temperature dependence. Although we cannot rule out the
built-in electric field picture, the agreement between the
theoretical explanations and the experimental results strongly
indicates that PTE may also be the origin of the PC in G/M and P-N
junction devices.

In summary, we have demonstrated that the PTE gives rise to the PC
generation at graphene interface field-effect transistors. The
temperature and power dependent results are in excellent agreement
with the PTE picture. Our work has potential impact for graphene
based opto-electronics, photo-thermocouple devices, and photovoltaic
applications.

This work is supported by the NSF through the Cornell Center for
Nanoscale Systems and Center for Materials Research, and by the
MARCO Focused Research Center on Materials, Structures, and Devices.
Device fabrication is performed at the Cornell Nano-Scale Science
and Technology Facility, funded by NSF.


\begin{thebibliography}{99}
\bibitem{note}A Strict comparison between the theory and data can be done if both spatial
and gate dependence of $R$ and $\Delta T$ are available.
\end{thebibliography}
\end{document}